\documentstyle[amssymb,12pt,thmsa,sw20lart]{article}
\input{tcilatex}
\begin{document}

\title{{\bf Supersymmetric structure of the induced }$W${\bf \ gravities}}
\author{Jean-Pierre Ader, Franck Biet, Yves Noirot \\
%EndAName
CPTM, CNRS-URA 1537, Universit\'{e} de Bordeaux I,\\
rue du Solarium, \\
F-33174 Gradignan Cedex, France.}
\maketitle

\begin{abstract}
We derive the supersymmetric structure present in $W$-gravities which has
been already observed in various contexts as Yang-Mills theory, topological
field theories, bosonic string and chiral $W_{3}$-gravity. This derivation
which is made in the geometrical framework of Zucchini, necessitates the
introduction of an appropriate new basis of variables which replace the
canonical fields and their derivatives. This construction is used, in the $%
W_{2}$-case, to deduce from the Chern-Simons action the Wess-Zumino-Polyakov
action.
\end{abstract}

\section{Introduction}

In this article we show that the supersymmetric structure found in
Chern-Simons theory quantized in the Landau gauge \cite{DGS} (so-called
since the anticommuting relations of the generators describe a super algebra
of the Wess-Zumino type) is also present in the induced $W_{n}$-gravities.
These theories \cite{ZAM} are higher spin generalizations of 2-dim gravity
whose symmetries are the classical $W_{n}$-algebras, where $n$ indicates the
highest spin of the currents involved. The geometrical description\cite{zuc}
is based on a straightforward generalization of the notion of projective
coordinates. It reproduces the results obtained in the more conventional
approaches based on $sl(n,{\em R})$ algebra \cite{P,BFK,DHR} . The reference
system of complex coordinates $(z,\overline{z})$ corresponds to a complex
structure defined on the connected, 2-topological manifold on which we are
working. Then the underlying models of the $W$-gravities are described by a
connection ${\cal A}=\Omega dz+\Omega ^{*}d\overline{z}$, with
zero-curvature condition

\begin{equation}
{\cal F}=d{\cal A-AA}=0  \label{1}
\end{equation}
. The pair of matrices $\left( \Omega ,\Omega ^{*}\right) $ contains
respectively the currents and the gauge fields of the theory which, in the
Virasoro case $n=2$, correspond respectively to the spin $2$ stress-energy
tensor $\rho _{zz}$ and the Beltrami coefficient $\mu _{\overline{z}}^{z}$ .
The fields involved here are smooth functions of the holomorphic coordinates 
$(z,\overline{z})$. They are built through a basic object, an unimodular
matrix $W$ (we denote by $(\partial ,\overline{\partial })$ the partial
derivatives with respect to $(z,\overline{z})$), 
\begin{equation}
\Omega =\partial WW^{-1}\;\;\;\;;\;\;\;\;\;\Omega ^{*}=\overline{\partial }%
WW^{-1}  \label{omw}
\end{equation}

The condition (\ref{1}) becomes\footnote{%
We adopt the convention that $\left[ A,B\right] $ stands for the
anticommutator if both $A$ and $B$ are grassmannian, else it is a commutator.%
}

\begin{equation}
\overline{\partial }\Omega -\partial \Omega ^{*}+\left[ \Omega ,\Omega
^{*}\right] =0  \label{(holo)}
\end{equation}
allowing us to determine the elements of $\Omega ^{*}$ in terms of the
elements of $\Omega $ and to give the holomorphy conditions obeyed by the
currents which are, in fact, the Ward identities of the theory. The main
advantage of \cite{zuc} is to derive easily the off-shell nilpotent BRST
algebra which expresses the invariance of the theory.

\begin{equation}
s{\cal A}=-dC+\left[ {\cal A},C\right] \,\,\,\,\,\,\,;\,\,\,\,\,sC=CC
\label{BRS1}
\end{equation}
where the ghost matrix $C$ is traced from $\Omega ^{*}$ by substituting the
gauge fields by the ghost fields. These laws are formally the BRST
transformations of the Yang-Mills (Y-M) connection ($d$ being the usual
derivative operator) and of the Faddeev-Popov ghost. The $W_{n}$-anomaly has
been obtained in this framework\cite{zuc,AAC} . After a general discussion
of the supersymmetric structure, we give explicit results in the case of $%
W_{2}$ and $W_{3}$-theories. Then a link with the group properties
underlying the formalism is made, allowing in principle a general
explicitation for $W_{n}.$ Some general remarks on the global properties of
this formulation are given. Finally the formalism is used to deduce from the
Chern-Simons action the Wess-Zumino-Polyakov action.

\section{Supersymmetric structure}

A new and simple way of solving descent equations has been recently
presented in ref.\cite{SS} where the case of the Y-M theory has been
treated. The method relies on the introduction of an operator $\Delta $
which allows one to decompose the exterior derivative $d$ as a BRST
commutator 
\begin{equation}
d=-[s,\Delta ].  \label{dh}
\end{equation}

The closure of the algebra among $d,s,$and $\Delta $ requires, in addition,
the introduction of a nilpotent operator $G$ such that the following
relations are obeyed

\begin{equation}
\lbrack d,\Delta ]=2G\,\,\,\,;\,\,\,\,\left[ d,G\right]
=0\,\,\,\,;\,\,\,\,\left[ s,G\right] =0\,\,\,\,;\,\,\,\,[\Delta ,G]=0
\label{algebre}
\end{equation}
This algebraic structure has already been found in topological field
theories such as the BF system \cite{BF}, the cohomological Witten's models%
\cite{CM} or the topological Yang-Mills theory\cite{TYM} , in bosonic string
theory in the Beltrami parametrization \cite{WSS} and in chiral $W_{3}$%
-gravity \cite{CQS} . In fact as we shall see now it appears also as a
general characteristic of $W_{n}$-induced gravities, following at once from
the parallelism with Y-M theory.

As in the Y-M case we define the linear operators of even and odd degrees
respectively

\begin{equation}
\Delta ={\cal A}\frac{\delta }{\delta C}+(2{\cal AA}-d{\cal A})\frac{\delta 
}{\delta (dC)}.  \label{del}
\end{equation}
\begin{equation}
G=(d{\cal A-AA)}\frac{\delta }{\delta C}+d({\cal AA})\frac{\delta }{\delta
(dC)}  \label{g}
\end{equation}
and we choose as independent variables the set (${\cal A},\;d{\cal A}%
,\;C,\;dC$). Note that the derivative of the connection ${\cal A}$ is taken
as a variable. On the local matrix space thus defined, the BRST operator $s$
and the exterior derivative $d$ act as ordinary differential operators.
Explicitly they read

\begin{eqnarray}
s &=&(-dC+[{\cal A},C])\frac{\delta }{\delta {\cal A}}+C^{2}\frac{\delta }{%
\delta C}-[dC,C]\frac{\delta }{\delta (dC)}-d[C,{\cal A}]\frac{\delta }{%
\delta (d{\cal A)}},  \label{ss1} \\
d &=&dC\frac{\delta }{\delta C}+d{\cal A}\frac{\delta }{\delta ({\cal A)}}\,
\label{dd3}
\end{eqnarray}
\vspace{0in}where the operator $[...]\delta /\delta \Phi $ replaces the $%
\Phi $ field by the expression $[...]$ by the expression in the right hand
side, and obeys the usual rules of derivatives in grassmannian space.

In the case of bilinear conformal theories, the system of descent equations
relates the cocycles of the cohomology in the following way

\begin{equation}
sT_{2}^{1}+dT_{1}^{2}=0\,\,\,;\,\,sT_{1}^{2}+dT_{0}^{3}=0\,\,\,;\,\,\,s%
\,T_{0}^{3}=0  \label{tour}
\end{equation}
where the lower index denotes the form degree and the upper index the ghost
number.

Now starting from the trace of the $C$ monomial of degree three, $%
T_{0}^{3}=Tr(CCC),$ the operator $\Delta $ generates the tower of descent
equations by giving the following solution for the cocycles

\begin{eqnarray}
T_{1}^{2} &=&\Delta T_{0}^{3}=3Tr(CC{\cal A})  \label{d4} \\
T_{2}^{1} &=&\frac{1}{2}\Delta \Delta T_{0}^{3}=3Tr(C{\cal AA})  \label{d6}
\end{eqnarray}
To get the last relation we have used the condition (\ref{1}). $T_{2}^{1}$
appears as a possible candidate for the non-integrated anomaly which
satisfies the Wess-Zumino consistency condition. Obviously this is in
agreement with the results\cite{AAC} obtained in the usual way, through the
Chern-Weil polynomial.

\section{The $n=2$ example}

In the simplest case of the $W_{2}$-model the connection ${\cal A}$ reads%
\cite{P,BFK}

\begin{equation}
{\cal A}=\left( 
\begin{array}{cc}
-\frac{1}{2}\partial \mu _{\bar{z}}^{z}d\overline{z} & \varkappa \smallskip
\\ 
-\frac{1}{2}\partial ^{2}\mu _{\bar{z}}^{z}d\overline{z}+\rho _{zz}ds & 
\frac{1}{2}\partial \mu _{\bar{z}}^{z}d\overline{z}
\end{array}
\right) \text{ }.  \label{os2}
\end{equation}
where $\varkappa $\smallskip $=dz+\mu _{\bar{z}}^{z}d\bar{z}$ . The set of
fields appearing in (\ref{os2}) are the Beltrami coefficient $\mu _{\bar{z}%
}^{z}$ ($\left| \mu _{\bar{z}}^{z}\right| \leqslant 1$) and the projective
connection $\rho _{zz}$ . The ghost matrix is obtained from the $\Omega ^{*}$
matrix by substituting for the Beltrami coefficient $\mu _{\bar{z}}^{z}$ and
the form degree the ghost field $c^{z}$ and the ghost degree respectively

\begin{equation}
C=\left( 
\begin{array}{cc}
-\frac{1}{2}\partial c^{z} & c^{z} \\ 
-\frac{1}{2}\partial ^{2}c^{z}+\rho _{zz}c^{z} & \frac{1}{2}\partial c^{z}
\end{array}
\right) .  \label{H}
\end{equation}

It is worthwhile to note that the differential form $\varkappa $\smallskip
appears in the well-known local relation involving the holomorphic
coordinates ($Z,\overline{Z}$) , corresponding to the structure parametrized
by $\mu _{\bar{z}}^{z}$%
\[
dZ=\partial Z(dz+\mu _{\bar{z}}^{z}d\bar{z})=\varkappa \smallskip \partial Z 
\]

Now the fields and their derivatives have to be considered as independent
variables. An adequate set of variables $\left\{
a_{1},a_{2},a_{3};c_{1},c_{2},c_{3}\right\} $ is dictated by the expression
of the matrices (\ref{os2},\ref{H}) and is given by

\begin{equation}
\begin{array}{lll}
a_{1}=\varkappa \smallskip 
\begin{array}{l}
;
\end{array}
& a_{2}=\partial \mu _{\bar{z}}^{z}d\bar{z} 
\begin{array}{l}
;
\end{array}
& a_{3}=\rho _{zz}\varkappa \smallskip -\frac{1}{2}\partial ^{2}\mu _{\bar{z}%
}^{z}d\bar{z} 
\begin{array}{l}
;
\end{array}
\end{array}
\label{cha}
\end{equation}

\begin{equation}
\begin{array}{lll}
c_{1}=c^{z} 
\begin{array}{l}
;
\end{array}
& c_{2}=\partial c^{z} 
\begin{array}{l}
;
\end{array}
& c_{3}=\rho _{zz}c^{z}-\frac{1}{2}\partial ^{2}c^{z} 
\begin{array}{l}
;
\end{array}
\end{array}
\label{chc}
\end{equation}

On the local space defined by these fields, the BRST operator $s$ and the
exterior derivative $d$ act as ordinary differential operators and are given
by

\[
d=\dsum\limits_{i=1}^{3}\left( dc_{i}\frac{\delta }{\delta c_{i}}+da_{i}%
\frac{\delta }{\delta a_{i}}\right) , 
\]

\[
s=\dsum\limits_{i=1}^{3}\left( sc_{i}\frac{\delta }{\delta c_{i}}+sa_{i}%
\frac{\delta }{\delta a_{i}}+sdc_{i}\frac{\delta }{\delta dc_{i}}+sda_{i}%
\frac{\delta }{\delta da_{i}}\right) 
\]
$.$

Note that the dimensional constraints coming from matching conformal indices
and the dimension of the matrices fix the number of higher order field
derivatives to two. The explicit forms of the BRST transformations of these
new fields are easily deduced from the matrix laws$\,\,(\ref{BRS1})\quad $

\begin{equation}
sa_{1}=-dc_{1}+a_{1}c_{2}+c_{1}a_{2},sa_{2}=-dc_{2}+2a_{3}c_{1}+2c_{3}a_{1},
\label{sa11}
\end{equation}

\begin{equation}
sa_{3}=-dc_{3}+a_{2}c_{3}+c_{2}a_{3},  \label{sa12}
\end{equation}

\begin{equation}
sc_{1}=c_{1}c_{2},\qquad sc_{2}=2c_{3}c_{1},\qquad sc_{3}=c_{2}c_{3}.
\label{sc11}
\end{equation}
whereas the $\Delta $ operator (acting on the space $\{c_{i},dc_{i}\}$) is
given by

\begin{equation}
\Delta =\dsum\limits_{i=1}^{3}\left( a_{i}\frac{\delta }{\delta c_{i}}%
+\Delta dc_{i}\frac{\delta }{\delta dc_{i}}\right) ,  \label{delta}
\end{equation}
with $\qquad \Delta dc_{1}=2a_{1}a_{2}-da_{1}$ ; $\Delta
dc_{2}=-4a_{1}a_{3}-da_{2}$ ; $\Delta dc_{3}=2a_{2}a_{3}-da_{3}.$ These
equalities betray the intrinsic relation between the $\Delta $ operator and
the BRST operator $s$ in this framework, since the relation between these
equalities and (\ref{sa11}) is obvious.

Finally, the algebra closes on

\[
{\cal G}=\dsum\limits_{i=1}^{3}\left( {\cal G}c_{i}\frac{\delta }{\delta
c_{i}}+{\cal G}dc_{i}\frac{\delta }{\delta dc_{i}}\right) , 
\]

with $\qquad {\cal G}c_{1}=da_{1}-a_{1}a_{2}$ $;{\cal G}%
c_{2}=da_{2}+2a_{1}a_{3}$ ;$\ {\cal G}c_{3}=da_{3}-a_{2}a_{3};$

and $\qquad {\cal G}dc_{1}=d(a_{1}a_{2})$ $;$ ${\cal G}%
dc_{2}=-2d(a_{1}a_{3}) $ $;$ ${\cal G}dc_{3}=d(a_{2}a_{3}).$

\bigskip Concerning the local cohomology of the BRST operator, we get from $%
T_{0}^{3}=Tr(CCC)$ and (\ref{d4},\ref{d6}) the cocycles in the zero and one
form sectors with ghost numbers three and two respectively which are already
known\cite{WSS} and the non-integrated anomaly yielding the usual expression
of the diffeomorphism anomaly $T_{2}^{1}=-\frac{3}{2}(\partial c^{z}\partial
^{2}\mu _{\bar{z}}^{z}-\partial ^{2}c^{z}\partial \mu _{\bar{z}}^{z})dzd\bar{%
z},$ for the bosonic string in the Beltrami parametrization\cite{BBSS} .

\smallskip Up to now we have ignored the fact that the fields have to obey
eq.(\ref{(holo)}). In fact these constraints are incorporated in the
explicit expressions (\ref{cha},\ref{chc}) of the new fields as functions of
the canonical ones. However if we want to derive this cohomology by applying
straightforwardly the BRST operator to the $a$ fields we have to use
explicitely the zero curvature condition, namely

\begin{equation}
da_{1}=a_{1}a\,_{2}\,\,;\,\,\,da_{2}=2a_{3}a_{1}\,\,\,;\,\,da_{3}=a_{2}a_{3}
\label{zcc}
\end{equation}
Note that the above first two conditions are used to determine the elements
of $\Omega ^{*}$ in terms of the elements of $\Omega $ , whereas the third
condition is the Ward identity of the theory.

\smallskip

\section{The induced $W_{3}$-gravity}

\smallskip

Now we present the way to derive the set of independent amplitudes
corresponding to the $W_{3}$-algebra. We consider, for instance, the ghost
matrix $\left\{ C_{ij}\right\} $which in this case reads\cite{AAC}

\begin{equation}
C=\left( 
\begin{array}{ccc}
\begin{array}{c}
\begin{array}{c}
\frac{1}{6}\partial ^{2}c^{zz}-\frac{2}{3}c^{zz}\rho _{zz}-\partial c^{z}
\end{array}
\\ 
\,\,
\end{array}
& 
\begin{array}{c}
\begin{array}{c}
c^{z}-\frac{1}{2}\partial c^{zz}
\end{array}
\\ 
\,
\end{array}
& 
\begin{array}{c}
\begin{array}{c}
c^{zz}
\end{array}
\\ 
\,
\end{array}
\\ 
\begin{array}{c}
\begin{array}{c}
\partial C_{11}-\frac{1}{2}c^{zz}\partial \rho _{zz}
\end{array}
\\ 
\,
\end{array}
& 
\begin{array}{c}
\begin{array}{c}
-\frac{1}{3}\left( \partial ^{2}c^{zz}-c^{zz}\rho _{zz}\right)
\end{array}
\\ 
\,
\end{array}
& 
\begin{array}{c}
\begin{array}{c}
c^{z}+\frac{1}{2}\partial c^{zz}
\end{array}
\\ 
\,
\end{array}
\\ 
\begin{array}{c}
\partial C_{21}+\partial \left( c^{zz}\rho _{zzz}\right) +c^{z}\rho _{zzz}
\\ 
\;\;\;\;\;\,\,\,\,\,\,\,+\frac{1}{2}c^{z}\partial \rho _{zz}
\end{array}
& 
\begin{array}{c}
\frac{1}{2}\partial C_{22}-\partial ^{2}c^{z}+c^{z}\rho _{zz} \\ 
+c^{zz}\rho _{zzz}
\end{array}
& \partial C_{23}+C_{22}
\end{array}
\right)  \label{fantome}
\end{equation}

From this matrix we have to determine eight independent ghost fields, three
of them corresponding to the fields of the $W_{2}$-algebra (which start with
a linear term of the form $\partial ^{n}c^{z}$, $n=0,1,2$), while the five
remaining fields have the linear terms $\partial ^{n}c^{zz},n=0..4$
respectively. Indeed, the formulation of the $W_{3}$-gravity has to contain
the $W_{2}$ results as a by-product. For instance, in the expression of the $%
W_{3}$-anomaly appears the $W_{2}$-anomaly. However, as recently shown \cite
{LAZZ}, this extension is partially formal and not well understood; it does
not provide a true Beltrami differential since now the modulus of the
Beltrami coefficient $\mu _{\overline{z}}^{z}$ appearing in the $W_{3}$%
-formalism, is no more necessarily less than one.

The issues of the band matrix extracted from (\ref{fantome}) by considering
the main diagonal and the two adjacent ones allow the determination of six
fields, whereas the two remaining fields are given by the two remaining
issues $C_{13}$ and $C_{31}$ of the matrix.

\begin{equation}
C=\left( 
\begin{array}{ccc}
\begin{array}{c}
c_{2}^{2}-c_{1}^{1}
\end{array}
\, & 
\begin{array}{c}
c_{1}^{0}-c_{2}^{1}
\end{array}
\, & 
\begin{array}{c}
c_{2}^{0}
\end{array}
\\ 
\begin{array}{c}
c_{2}^{3}-c_{1}^{2}
\end{array}
& 
\begin{array}{c}
-2c_{2}^{2}
\end{array}
& 
\begin{array}{c}
c_{1}^{0}+c_{2}^{1}
\end{array}
\, \\ 
\begin{array}{c}
c_{2}^{4}
\end{array}
& 
\begin{array}{c}
-c_{2}^{3}-c_{1}^{2}
\end{array}
& 
\begin{array}{c}
c_{2}^{2}+c_{1}^{1}
\end{array}
\end{array}
\right)  \label{ff3}
\end{equation}

\medskip

In $c_{i}^{j}$ the indices refer to the linear term contained in its
expression: $j$ is the derivative power appearing in this term while $i$
means that the ghost on which this derivative acts concerns the $W_{2}$%
-algebra ($i=1$) or the $W_{3}$-algebra ($i=2$). For instance $c_{2}^{1}$
contains $\partial c^{zz}.$

The fields corresponding to the connection matrix $\left\{ {\cal A}%
_{ij}\right\} $ can be obtained in the same way as above. Starting with\cite
{AAC}

${\cal A=}\left( 
\begin{array}{ccc}
\begin{array}{c}
\left( \frac{1}{6}\partial ^{2}\mu _{\overline{z}}^{zz}-\frac{2}{3}\mu _{%
\overline{z}}^{zz}\rho _{zz}-\partial \mu _{\overline{z}}^{z}\right) d%
\overline{z} \\ 
\,
\end{array}
& 
\begin{array}{c}
\varkappa -\frac{1}{2}\partial \mu _{\overline{z}}^{zz}d\overline{z} \\ 
\,
\end{array}
& 
\begin{array}{c}
\begin{array}{c}
\mu _{\overline{z}}^{zz}d\overline{z}
\end{array}
\\ 
\,
\end{array}
\\ 
\begin{array}{c}
\partial {\cal A}_{11}-\frac{1}{2}\mu _{\overline{z}}^{zz}\partial \rho
_{zz}d\overline{z} \\ 
\,
\end{array}
& 
\begin{array}{c}
-\frac{1}{3}\left( \partial ^{2}\mu _{\overline{z}}^{zz}-\mu _{\overline{z}%
}^{zz}\rho _{zz}\right) d\overline{z} \\ 
\,
\end{array}
& 
\begin{array}{c}
\begin{array}{c}
\varkappa +\frac{1}{2}\partial \mu _{\overline{z}}^{zz}d\overline{z}
\end{array}
\\ 
\,
\end{array}
\\ 
\begin{array}{c}
\partial {\cal A}_{21}++\rho _{zzz}\varkappa \\ 
+\left( \partial \left( \mu _{\overline{z}}^{zz}\rho _{zzz}\right) +\frac{1}{%
2}\mu _{\overline{z}}^{z}\partial \rho _{zz}\right) d\overline{z}
\end{array}
& 
\begin{array}{c}
\frac{1}{2}\partial {\cal A}_{22}+\rho _{zz}\varkappa \\ 
-\left( \partial ^{2}\mu _{\overline{z}}^{z}-\mu _{\overline{z}}^{zz}\rho
_{zzz}\right) d\overline{z}
\end{array}
& \partial {\cal A}_{23}+{\cal A}_{22}
\end{array}
\right) ,$

and defining

\begin{equation}
{\cal A=}\left( 
\begin{array}{ccc}
\begin{array}{c}
a_{2}^{2}-a_{1}^{1}
\end{array}
\, & 
\begin{array}{c}
a_{1}^{0}-a_{2}^{1}
\end{array}
& 
\begin{array}{c}
a_{2}^{0}
\end{array}
\\ 
\begin{array}{c}
a_{2}^{3}-a_{1}^{2}
\end{array}
\, & 
\begin{array}{c}
-2a_{2}^{2}
\end{array}
\, & 
\begin{array}{c}
a_{1}^{0}+a_{2}^{1}
\end{array}
\\ 
\begin{array}{c}
a_{2}^{4}
\end{array}
& 
\begin{array}{c}
-a_{2}^{3}-a_{1}^{2}
\end{array}
& 
\begin{array}{c}
a_{2}^{2}+a_{1}^{1}
\end{array}
\end{array}
\right)  \label{ff4}
\end{equation}

\smallskip

it is straightforward to read off the explicit expressions of $a_{i}$. The
indices have the same meaning as before, the upper indices refering now to
the linear terms of the $\mu $'s fields\footnote{%
A careful reader would note that the expression of the $a_{1}^{2}$ field is
different from the expression of the $a_{3}$ field given previously in the $%
W_{2}$-formalism since $a_{1}^{2}=-\partial ^{2}\mu _{\overline{z}}^{z}d%
\overline{z}+\frac{1}{2}\rho _{zz}\varkappa $. This is due to a trivial
redefinition of the fields in the $W_{3}$-framework.}. For instance the
expression of $a_{2}^{1}$ contains the term $\partial \mu _{\overline{z}%
}^{zz}$. The upper index corresponds to the power of the derivative whereas
the index $2$ indicates that $\mu _{\overline{z}}^{zz}$ is a field of the $%
W_{3}$ algebra.

Having identified the basic fields in the connection and ghost field sectors
we give now their BRST transformations.

\begin{center}
$
\begin{array}{ccc}
\begin{array}{c}
sc_{1}^{0}=3c_{2}^{1}c_{2}^{2}+c_{2}^{3}c_{2}^{0}+c_{1}^{0}c_{1}^{1} \\ 
\,
\end{array}
& 
\begin{array}{c}
sc_{1}^{1}=c_{2}^{4}c_{2}^{0}+c_{1}^{0}c_{1}^{2}+c_{2}^{1}c_{2}^{3} \\ 
\,
\end{array}
& 
\begin{array}{c}
sc_{1}^{2}=3c_{2}^{2}c_{2}^{3}+c_{2}^{4}c_{2}^{1}+c_{1}^{1}c_{1}^{2} \\ 
\,
\end{array}
\\ 
\begin{array}{c}
sc_{2}^{0}=2\left( c_{2}^{0}c_{1}^{1}+c_{1}^{0}c_{2}^{1}\right) \\ 
\,
\end{array}
& 
\begin{array}{c}
sc_{2}^{1}=3c_{1}^{0}c_{2}^{2}+c_{2}^{0}c_{1}^{2}+c_{2}^{1}c_{1}^{1\,} \\ 
\,
\end{array}
& 
\begin{array}{c}
sc_{2}^{2}=c_{1}^{0}c_{2}^{3}+c_{2}^{1}c_{1}^{2} \\ 
\,
\end{array}
\end{array}
$

$
\begin{array}{ccc}
\begin{array}{c}
sc_{2}^{3}=3c_{2}^{2}c_{1}^{2}+c_{1}^{1}c_{2}^{3}+c_{1}^{0}c_{2}^{4} \\ 
\,
\end{array}
& \, & \,\,\,\,\,\,\,\,\,\,\,\, 
\begin{array}{c}
sc_{2}^{4}=2\left( c_{2}^{3}c_{1}^{2}+c_{1}^{1}c_{2}^{4}\right) \\ 
\,
\end{array}
\end{array}
$

$
\begin{array}{cc}
\begin{array}{c}
sa_{1}^{0}=-dc_{1}^{0}+3\left[ a_{2}^{1},c_{2}^{2}\right] +\left[
a_{1}^{0},c_{1}^{1}\right] +\left[ a_{2}^{3},c_{2}^{0}\right] \\ 
\,
\end{array}
& 
\begin{array}{c}
sa_{1}^{1}=-dc_{1}^{1}+\left[ a_{2}^{1},c_{2}^{3}\right] +\left[
a_{1}^{0},c_{1}^{2}\right] +\left[ a_{2}^{4},c_{2}^{0}\right] \\ 
\,
\end{array}
\end{array}
$

$
\begin{array}{c}
sa_{1}^{2}=-dc_{1}^{2}+3\left[ a_{2}^{2},c_{2}^{3}\right] +\left[
a_{1}^{1},c_{1}^{2}\right] +\left[ a_{2}^{4},c_{2}^{1}\right] \\ 
\,
\end{array}
$

$
\begin{array}{cc}
\begin{array}{c}
sa_{2}^{0}=-dc_{2}^{0}+2\left[ a_{2}^{0},c_{1}^{1}\right] +2\left[
a_{1}^{0},c_{2}^{1}\right] \\ 
\,
\end{array}
& 
\begin{array}{c}
sa_{2}^{1}=-dc_{2}^{1}+3\left[ a_{1}^{0},c_{2}^{2}\right] +\left[
a_{2}^{1},c_{1}^{1}\right] +\left[ a_{2}^{0},c_{1}^{2}\right] \\ 
\,
\end{array}
\\ 
\begin{array}{c}
sa_{2}^{2}=-dc_{2}^{2}+\left[ a_{1}^{0},c_{2}^{3}\right] +\left[
a_{2}^{1},c_{1}^{2}\right] \\ 
\,
\end{array}
& 
\begin{array}{c}
sa_{2}^{3}=-dc_{2}^{3}+3\left[ a_{2}^{2},c_{1}^{2}\right] +\left[
a_{1}^{1},c_{2}^{3}\right] +\left[ a_{1}^{0},c_{2}^{4}\right] \\ 
\,
\end{array}
\end{array}
$

$
\begin{array}{c}
sa_{2}^{4}=-dc_{2}^{4}+2\left[ a_{1}^{1},c_{2}^{4}\right] +2\left[
a_{2}^{3},c_{1}^{2}\right] . \\ 
\,
\end{array}
$
\end{center}

The brackets are defined by $\left[ a_{j}^{i},c_{l}^{k}\right]
=a_{j}^{i}c_{l}^{k}+c_{j}^{i}a_{l}^{k}$ and obey $\left[
a_{j}^{i},c_{l}^{k}\right] =\left[ c_{j}^{i},a_{l}^{k}\right] =-\left[
c_{l}^{k},a_{j}^{i}\right] .$ The $\Delta $ operator has the same form as in
equation (\ref{delta}) with $\Delta c_{i}=a_{i}.$ The expression of $\Delta
dc_{i}$ is simply deduced from the BRST transformations above by replacing
the $c_{i}^{j}$'s by the $a_{i}^{j}$'s.

Let us proceed to give the construction of the anomaly. Starting from the
cocycle

\begin{equation}
T_{0}^{3}=tr(C^{3})=6(-3c_{1}^{0}c_{2}^{2}c_{2}^{3}+3c_{2}^{1}c_{1}^{2}c_{2}^{2}+c_{1}^{0}c_{1}^{2}c_{1}^{1}+c_{1}^{1}c_{2}^{1}c_{2}^{3}+c_{1}^{0}c_{2}^{1}c_{2}^{4}-c_{2}^{0}c_{1}^{2}c_{2}^{3}+c_{2}^{0}c_{1}^{1}c_{2}^{4}),
\label{brs2}
\end{equation}

\smallskip the cocycles of the descent equations are obtained by the action
of the operator $\Delta $ and the following expression of the anomaly is
easily deduced 
\[
T_{2}^{1}=6(-3a_{1}^{0}a_{2}^{2}c_{2}^{3}+3a_{2}^{1}a_{1}^{2}c_{2}^{2}+a_{1}^{0}a_{1}^{2}c_{1}^{1}+a_{1}^{0}a_{2}^{1}c_{2}^{4}-a_{2}^{0}a_{1}^{2}c_{2}^{3}-3c_{1}^{0}a_{2}^{2}a_{2}^{3} 
\]
\[
+3c_{2}^{1}a_{1}^{2}a_{2}^{2}+c_{1}^{0}a_{1}^{2}a_{1}^{1}+c_{1}^{1}a_{2}^{1}a_{2}^{3}+c_{1}^{0}a_{2}^{1}a_{2}^{4}-c_{2}^{0}a_{1}^{2}a_{2}^{3}+c_{2}^{0}a_{1}^{1}a_{2}^{4}-3a_{1}^{0}c_{2}^{2}a_{2}^{3} 
\]
\[
+3a_{2}^{1}c_{1}^{2}a_{2}^{2}+a_{1}^{0}c_{1}^{2}a_{1}^{1}+a_{1}^{1}c_{2}^{1}a_{2}^{3}+a_{1}^{0}c_{2}^{1}a_{2}^{4}-a_{2}^{0}c_{1}^{2}a_{2}^{3}+a_{2}^{0}c_{1}^{1}a_{2}^{4}). 
\]

The final form of this quantity in terms of the basic fields $\mu _{%
\overline{z}}^{z},$ $\rho _{zz},\mu _{\overline{z}}^{zz},\rho _{zzz}$ is
straightforwardly available from the explicit expressions of the $a$ and $c$
fields and is not given here since it is already known \cite{AAC,OSSN} . The
only point we want to discuss is that the terms $%
6(a_{1}^{0}a_{1}^{2}c_{1}^{1}+a_{1}^{0}c_{1}^{2}a_{1}^{1})$ and $%
-6(3a_{1}^{0}a_{2}^{2}c_{2}^{3}+3a_{1}^{0}c_{2}^{2}a_{2}^{3}-a_{1}^{0}a_{2}^{1}c_{2}^{4}-a_{1}^{0}c_{2}^{1}a_{2}^{4}) 
$ contains respectively the leading terms $\partial ^{2}\mu _{\overline{z}%
}^{z}\partial c^{z}-\partial ^{2}c^{z}\partial \mu _{\overline{z}}^{z}$ and $%
\partial ^{2}\mu _{\overline{z}}^{zz}\partial ^{3}c^{zz}-\partial
^{3}c^{zz}\partial ^{2}\mu _{\overline{z}}^{zz}$ ( called universal
anomalies by Hull \cite{H1}).

\smallskip

\section{The general formulation}

\smallskip

From the examples of the $W_{2}$ and $W_{3}$ -models we can draw some
general lessons. There are $n^{2}-1$ fields (and $n^{2}-1$ ghosts) necessary
to describe the $W_{n}$-model. They are decomposed in the following way : 
\begin{equation}
n^{2}-1=\sum\limits_{i=2}^{n}(2i-1)  \label{decomp}
\end{equation}
where each term in the sum corresponds to the subset of fields describing
the $W_{i}$-model. This express the fact that the $W_{n}$-algebra contains
the nested set of subalgebras $W_{k}$ , $k=2,..,n-1$ : $W_{2}\subset
W_{3}\subset ..\subset W_{n-1}\subset W_{n}$, where the inclusion symbol
means that the formulation of $W_{i}$ can be obtained from $W_{i+1}$ by
setting to zero the fields occuring at the level $i+1$. Finally the fields
of the $W_{n}$-model are deduced from the $2(n-1)$ non-principal diagonals
following the decomposition $2\sum\limits_{i=1}^{n-1}i$ , the main diagonal
giving the $n-1$ remaining fields. We note that the constraints imposed by
the conformal indices and the size of the matrices, building blocks of the $%
W_{n}$-model, imply that the degree of derivatives appearing in the
expressions of the fields must be $2n-2$ at most.

More importantly, the BRST algebra of the ghost fields of the $W_{n}$-models
expressed in terms of the new fields (see eq.(\ref{sc11}) for the example of 
$W_{2}$ and the expressions given in sect.4. for $W_{3}$ ) reflects the
group symmetry $sl(n,{\em R})$. To prove this let us remember the link
between a Lie algebra and an antiderivation operator. Let ${\cal G}$ be a
vector space of dimension $N$ with a basis ($T_{\alpha ,}$ $\alpha =1,N$ ).
The corresponding Lie algebra structure is defined by writing a commutator
of two generators of ${\cal G}$ as $[T_{\alpha },T_{\beta }]=f_{\alpha \beta
}^{\gamma }T_{\gamma }$ , where $f_{\alpha \beta }^{\gamma }$ are the
antisymmetric structure constants. Moreover the commutators satisfy the
Jacobi identity $[[T_{\alpha },T_{\beta }],T_{\gamma }]+cyclic$ $%
permutations=0$ . The Lie algebra may be also defined in the dual space $%
{\cal G}^{*}$ of ${\cal G}$ (with the wedge product $\wedge $ between its
elements). In a dual basis ($C^{\alpha },$ $\alpha =1,...,N$ ) the
antiderivation $s$ of degree $1$ is defined by

\begin{equation}
sC^{\alpha }=\frac{1}{2}f_{\beta \gamma }^{\alpha }C^{\beta }\wedge
C^{\gamma }  \label{der}
\end{equation}
It is easy to verify that the nilpotency condition of $s$ results from the
Jacobi identity and that the algebra (\ref{sc11}) corresponds to the Lie
brackets satisfied by the generators of the groups $sl(2,{\em R})$ when they
are identified with (\ref{der}). The transformations of the gauge fields are
those of a Yang-Mills theory of the $sl(2,{\em R})$ group: 
\[
sa^{\alpha }=-dC^{\alpha }+\frac{1}{2}f_{\beta \gamma }^{\alpha }a^{\beta
}C^{\gamma } 
\]

The generalization to higher $W_{n}$-models requires to take into account
the nested structure mentioned before. If $\underline{2j+1}_{2}$ denotes the 
$(2j+1)-$dimensional irreducible representation of $sl(2,{\em R})$, the $%
(n^{2}-1)-$dimensional adjoint representation of $sl(n,{\em R})$, $%
\underline{ad}_{n}$ has the branching rule 
\[
\underline{ad}_{n}\backsimeq \underline{3}_{2}\oplus \underline{5}_{2}\oplus
...\oplus \underline{2n-1}_{2} 
\]

From this follows immediately the decomposition (\ref{decomp}). Without
going into details\footnote{%
The technology of $sl$ representations has been recently reviewed in \cite
{BT}. We borrow the group notations and conventions to these publications}
we can give the corresponding basis of $sl(3,{\em R})$%
\[
J^{a}t_{a}=\left( 
\begin{array}{ccc}
\begin{array}{c}
\frac{J^{4}}{6}+\frac{J^{5}}{2}
\end{array}
\, & 
\begin{array}{c}
\frac{J^{6}}{2}+\frac{J^{7}}{2}
\end{array}
& 
\begin{array}{c}
J^{8}
\end{array}
\\ 
\begin{array}{c}
\frac{J^{2}}{2}+\frac{J^{3}}{2}
\end{array}
\, & 
\begin{array}{c}
-\frac{J^{4}}{3}
\end{array}
\, & 
\begin{array}{c}
\begin{array}{c}
-\frac{J^{6}}{2}+\frac{J^{7}}{2}
\end{array}
\end{array}
\\ 
\begin{array}{c}
J^{1}
\end{array}
& 
\begin{array}{c}
\frac{J^{2}}{2}-\frac{J^{3}}{2}
\end{array}
& 
\begin{array}{c}
\begin{array}{c}
\frac{J^{4}}{6}-\frac{J^{5}}{2}
\end{array}
\end{array}
\end{array}
\right) 
\]
where the $sl(2,{\em R})$ subalgebra is given by $J^{2},J^{5}$ and $J^{7}$.
The comparison with (\ref{ff3}) and (\ref{ff4}) is obvious. Finally in order
to illustrate the construction in a less simple case we give the example of $%
sl(4,{\em R}):$%
\[
J^{a}t_{a}=\left( 
\begin{array}{cccc}
\frac{J^{7}}{2}+J^{8}+J^{9} & J^{5}+J^{6} & J^{2}+J^{3} & J^{1} \\ 
J^{10}+J^{11} & \frac{J^{7}}{2}-J^{8}-J^{9} & J^{4} & J^{2}-J^{3} \\ 
J^{12}+J^{13} & J^{14} & -\frac{J^{7}}{2}+J^{8}-J^{9} & J^{5}-J^{6} \\ 
J^{15} & J^{12}-J^{13} & J^{10}-J^{11} & -\frac{J^{7}}{2}-J^{8}+J^{9}
\end{array}
\right) 
\]
where now the $sl(2,{\em R})$ subalgebra is given by $J^{2},J^{7}$ and $%
J^{12}$ and the $sl(3,{\em R})$ subalgebra is given by $%
J^{3},J^{6},J^{8},J^{10}$ and $J^{13}.$

\section{Global aspects of the formulation}

\smallskip

Clearly the formalism described up to now, is valid only on the plane and on
the sphere. When considering a Riemann surface of higher genus a global
formulation of the anomaly and of the cocycles linked to it by the
cohomology of the BRST operator is possible\cite{BG,AAC}. However the field
used to render the expressions valid on any local coordinate chart is the $%
\rho _{zz}$ field appearing in (\ref{os2}) and (\ref{H}). Indeed this field
transforms with the Schwarzian derivative under a conformal change of
coordinates but is not a true projective connection since it is not locally
holomorphic ($\overline{\partial }\rho _{zz}\neq 0$). In fact it obeys to
the holomorphic condition (for the $W_{2}$-model) 
\begin{equation}
(\overline{\partial }-\mu _{\overline{z}}^{z}\partial -2\partial \mu _{%
\overline{z}}^{z})\rho _{zz}=-\frac{1}{2}\partial ^{3}\mu _{\bar{z}}^{z}
\label{WA}
\end{equation}

which is formally the anomalous Ward identity of the induced 2-dim.
conformal gravity. This results in a serious drawback of this formulation
since the underlying field theory is ill defined, the holomorphic condition (%
\ref{WA}) linking in a non-local way the $\mu _{\overline{z}}^{z}$ and the $%
\rho _{zz}$ fields. The situation is even worse for the $W_{3}$-model since
then the $\rho _{zz}$ field appears explicitly in the BRST transformations
of the $c^{z}$ and $\mu _{\overline{z}}^{z}$ whereas the known expressions
of the non-integrated anomaly \cite{AAC,OSSN} display terms containing the $%
\rho _{zz}$ and $\rho _{zzz}$ fields. In fact this defect is inherent to any
calculation starting from the zero curvature condition (\ref{(holo)}) and
being self-contained in the sense that the field used to glue the
expressions from different charts is provided by the framework itself.

In \cite{LZ} a different point of view was adopted. Working with a
holomorphic projective connection and thus avoiding to introduce
non-locality in the theory, Lazzarini and Stora derived the form of the
Virasoro Ward identity (\ref{WA}) on arbitrary Riemann surfaces.
Unfortunately their work cannot be carried through for the $W_{3}$-model.
Indeed replacing the $\rho $ fields appearing in the $\Omega $ matrix by
holomorphic fields (and thus BRST\ inerts) obviously breaks the BRST
invariance of the integrated (or non-integrated) anomaly as well as the
nilpotency of the BRST transformations of $c^{z}$ and $\mu _{\overline{z}%
}^{z}$ . The systematic way to formulate the theory provided by the
geometrical framework of \cite{zuc} or given by some zero curvature
condition \cite{BFK,DHR,OSSN,BG} is lost.

\smallskip

\section{The Wess-Zumino-Polyakov action}

\smallskip In this section we derive the Wess-Zumino-Polyakov action from
the Chern-Simons action for the $W_{2}$-model. Following the formalism at
work in Y-M theory\cite{ENSS} the Chern-Simons action is given by 
\[
S=\frac{k}{4\pi }\int_{Y}Tr\left( \widetilde{{\cal A}}\widetilde{d}%
\widetilde{{\cal A}}-\frac{2}{3}\widetilde{{\cal A}}\widetilde{{\cal A}}%
\widetilde{{\cal A}}\right) 
\]
This action is defined on a 3-dimensional manifold $Y$ whose boundary is the
two dimensional space $(z,\overline{z})$. More precisely we assume for $Y$ a
''space-time'' splitting of the form $\Sigma \times R$ where $\Sigma $ is a
Riemann surface (eventually with boundary). Then the exterior derivative $%
\widetilde{d}=dt\partial _{t}+d$ ($\partial _{t}\equiv \frac{\partial }{%
\partial t}$) and the matrix form $\widetilde{{\cal A}}=$ ${\cal A}_{t}+%
{\cal A}$ are decomposed into a ''time'' component and the usual space
components $(z,\overline{z})$. The action above becomes 
\begin{equation}
S=-\frac{k}{4\pi }\int_{Y}Tr\left( {\cal A}\partial _{t}{\cal A}dt\right) +%
\frac{k}{2\pi }\int_{Y}Tr\left( {\cal A}_{t}(d{\cal A}-{\cal AA)}\right)
\label{csa}
\end{equation}

Here ${\cal A}_{t}$ can be viewed as a Lagrange multiplier enforcing the
constraints (\ref{1}) in the space direction. An effective action can then
be derived by substituting the expression of ${\cal A}$ in terms of the
variables \{$a_{i}$\}into (\ref{csa}). 
\begin{equation}
S=\int_{Y}\left( a_{1\overline{z}}\partial _{t}a_{3z}-a_{3z}\partial _{t}a_{1%
\overline{z}}+a_{3\overline{z}}\partial _{t}a_{1z}-a_{1z}\partial _{t}a_{3%
\overline{z}}+\frac{1}{2}a_{2\overline{z}}\partial _{t}a_{2z}-\frac{1}{2}%
a_{2z}\partial _{t}a_{2\overline{z}}\right)  \label{awzw}
\end{equation}
The equations of motion are the zero curvature conditions (\ref{zcc}). Now
let $\Sigma =D$ be a disk; these constraints are automatically satisfied
when the gauge matrix field ${\cal A}$ is locally expressed as a pure gauge: 
\[
{\cal A}=dWW^{-1}=\left( 
\begin{array}{cc}
0 & 1 \\ 
\partial ^{2}g\partial h-\partial ^{2}h\partial g & 0
\end{array}
\right) dz+\left( 
\begin{array}{cc}
\overline{\partial }g\partial h-\partial g\overline{\partial }h & \overline{%
\partial }hg-h\overline{\partial }g \\ 
\partial \overline{\partial }g\partial h-\partial \overline{\partial }%
h\partial g & -\overline{\partial }g\partial h+\partial g\overline{\partial }%
h
\end{array}
\right) d\overline{z}\;\; 
\]
where $W=\left( 
\begin{array}{cc}
g & h \\ 
\partial g & \partial h
\end{array}
\right) $ satisfies to 
\begin{equation}
g\partial h-h\partial g=1  \label{uni}
\end{equation}
The peculiar structure of the matrix $W$, a Wronskian structure, determines
the form of the matrix $\Omega $ as a function of the two independent fields
($\mu _{\bar{z}}^{z},\rho _{zz}$). The choice of two (-$\frac{1}{2}$)
differentials $g=\frac{1}{\sqrt{\partial Z}}\,\,,\,\,h=\frac{Z}{\sqrt{%
\partial Z}}$ makes the link with the theory of two-dimensional Riemann
surfaces\cite{zuc,P}, since $Z$ is in fact, a generic coordinate of the
conformal structure ${\bf A}(\mu )$ associated to any Beltrami differential $%
\mu $ . It is a local solution of the Beltrami equation and $\,\,\,\rho
_{zz} $ is the Schwarzian derivative of $Z$.

The appearance of a third coordinate requires some explanation. It is
assumed that the matrix field $W(t,z,\overline{z})$ is defined on a
three-dimensional hemisphere whose boundary $t_{B}$ coincides with the
two-dimensional plane where the original theory is defined in such a way
that $W(t_{B},z,\overline{z})\equiv W(z,\overline{z})$. Now by following the
canonical formalism of Witten it is possible to relate the action (\ref{csa}%
) to the theory in two dimensions. Integrating by parts we get 
\begin{equation}
\begin{array}{c}
S=-\frac{k}{4\pi }\int dtd\overline{z}Tr(\overline{\partial }WW^{-1}\partial
_{t}WW^{-1})-\frac{k}{4\pi }\int dtdzTr(\partial WW^{-1}\partial _{t}WW^{-1})
\\ 
+\frac{k}{12\pi }\int_{Y}Tr(dWW^{-1})^{3}
\end{array}
\label{WZW}
\end{equation}
The change of variables from ${\cal A}$ to $W$ involves a unit Jacobian
since the measure $d\widetilde{{\cal A}}\delta ({\cal F})\equiv dW$. In this
way we have a path integral of the action (\ref{WZW}) which looks like the
chiral version of the Wess-Zumino-Witten path integral.

This action depends only on the boundary values of $W$ on the conformal
space $(z,\overline{z})$. The third term on the r.h.s of (\ref{WZW}) is the
generalization of the usually called Wess-Zumino term \cite{WZ}, and in fact
does not depend on the $t$ variable being a total derivative. To prove this
latter statement it is worthwhile to restore in its expression the full
symmetry with respect to the three variables $z,\overline{z}$ and $t$ . This
results in 
\[
\begin{array}{c}
\int_{Y}Tr(dWW^{-1})^{3}=\int_{Y}[(\partial \overline{\partial }g\partial
h-\partial \overline{\partial }h\partial g)(\partial _{t}g\partial
h-\partial g\partial _{t}h) \\ 
-(\partial \partial _{t}g\partial h-\partial \partial _{t}h\partial g)(%
\overline{\partial }g\partial h-\partial g\overline{\partial }h)+(\partial
^{2}g\partial h-\partial ^{2}h\partial g)(\overline{\partial }g\partial
_{t}h-\partial _{t}g\overline{\partial }h)]
\end{array}
\]
With the help of the condition (\ref{uni}) we see that this three
dimensional integral depends on boundary values only 
\[
\begin{array}{c}
\int_{Y}Tr(dWW^{-1})^{3}=\int_{Y}\partial _{t}[\ln g(\partial \overline{%
\partial }g\partial h-\overline{\partial }h\partial ^{2}g)]+\int_{Y}\partial
[\ln g(\partial ^{2}g\partial _{t}h-\partial h\partial \partial _{t}g)] \\ 
+\int_{Y}\overline{\partial }[\ln g(\partial \partial _{t}g\overline{%
\partial }h-\partial _{t}h\partial \overline{\partial }g)]
\end{array}
\]
The formal expression $\ln g$ which appears in the formula above has been
obtained by partial integration of $\frac{\partial g}{g}$. Now we assume
that $g$ and $h$ satisfy on the boundary of the space ($z$,$\overline{z}$)
the relation $h=\partial g$ (and $g\partial ^{2}g-(\partial g)^{2}=1$ which
preserves (\ref{uni})). The three dimensional integral is reduced to 
\[
\int_{Y}Tr(dWW^{-1})^{3}=\int dzd\overline{z}\ln g(\partial \overline{%
\partial }g\partial h-\overline{\partial }h\partial ^{2}g) 
\]
On the Riemann surface the result is (up to some integration by part) the
Wess-Zumino-Polyakov action 
\[
\int_{Y}Tr(dWW^{-1})^{3}=\int dzd\overline{z}\frac{\overline{\partial }Z}{%
\partial Z}\partial ^{2}\ln \partial Z 
\]
which solves the Ward identity. In particular this is a non-local functional
of the Beltrami coefficient obeying 
\[
\frac{\delta }{\delta \mu }\int dzd\overline{z}\frac{\overline{\partial }Z}{%
\partial Z}\partial ^{2}\ln \partial Z=2S(Z,z) 
\]
where $S$ is the Schwarzian derivative.

Now we explain how the above results are related to other calculations of
the Wess-Zumino-Polyakov action. In fact starting from \cite{OSSN,BG} 
\begin{equation}
\int dzd\overline{z}\frac{\overline{\partial }Z}{\partial Z}\partial ^{2}\ln
\partial Z=\int dzd\overline{z}Tr(\partial gg^{-1}\overline{\partial }%
gg^{-1})-\int dzd\overline{z}Tr(\Lambda g^{-1}\overline{\partial }%
g)+\int_{Y}Tr(dWW^{-1})^{3}  \label{boule}
\end{equation}
where $\Lambda $ is the constant matrix $\left( 
\begin{array}{cc}
0 & 1 \\ 
0 & 0
\end{array}
\right) $, the expressions in the r.h.s of (\ref{boule}) can be reexpressed
as 
\[
\begin{array}{c}
\int dzd\overline{z}\frac{\overline{\partial }Z}{\partial Z}\partial ^{2}\ln
\partial Z=\frac{1}{4}\int dzd\overline{z}(2\mu _{\bar{z}}^{z}\rho _{zz}-%
\frac{1}{2}\partial ^{2}\mu _{\bar{z}}^{z})-\frac{1}{2}\int dzd\overline{z}%
(\mu _{\bar{z}}^{z}\rho _{zz}-\frac{1}{2}\partial ^{2}\mu _{\bar{z}}^{z}) \\ 
+\int_{Y}Tr(dWW^{-1})^{3}
\end{array}
\]
Since the first two terms in the r.h.s reduce to a total derivative, we
obtain the desired result.

We have clarified how the Wess-Zumino-Witten action can be obtained in this
framework. We hope that this derivation will be generalizable to other
bi-dimensional conformal models.

{\bf Acknowledgments} : We would like to thank P.Minnaert and

R. Stora for interesting discussions.

\end{document}